\documentclass[12pt,article]{revtex4}
\usepackage{graphicx}
\usepackage{mathpazo}

\begin{document}

\title{Quantum Cylindrical Waves and Sigma Models}

\author{J. Fernando Barbero G.\thanks{jfbarbero@imaff.cfmac.csic.es}
and Guillermo A. Mena Marug\'an\thanks{mena@imaff.cfmac.csic.es}
\\{\sl I.M.A.F.F.,
Centro de F\'{\i}sica Miguel A. Catal\'{a}n, C.S.I.C., C/ Serrano
113bis/121, 28006 Madrid, Spain}\vspace*{.4cm}\\ Eduardo J. S.
Villase\~nor
\thanks{ejsanche@math.uc3m.es}\\{\sl Departamento de
Matem\'aticas, Escuela Polit\'ecnica Superior,}\\{\sl Universidad
Carlos III de Madrid, Avda. de la Universidad 30, 28911 Legan\'es,
Spain}\vspace*{.4cm}}

\date{December 15, 2003}

\begin{abstract}
We analyze cylindrical gravitational waves in vacuo with general
polarization and develop a viewpoint complementary to that
presented recently by Niedermaier showing that the auxiliary sigma
model associated with this family of waves is not renormalizable
in the standard perturbative sense.
\end{abstract}

\maketitle

\renewcommand{\thesection}{\arabic{section}}

\section{\label{Intro}Introduction}

Einstein-Rosen waves, i.e. linearly polarized cylindrical waves,
have been extensively studied in the literature, both from the
classical and quantum viewpoints \cite{Einstein, Kuchar, Ashtekar,
Madhavan, Ashtekar1, Niedermaier2}. They provide a non-trivial
reduction of General Relativity (GR) that retains some important
features of the full theory. In particular, they are described by
a genuine field theory, with one local degree of freedom. However,
one of the most distinctive features of GR, namely non-linearity,
is not fully realized in this system inasmuch as the reduced phase
space of the Einstein-Rosen waves is equivalent to that of an
auxiliary linear theory, without appealing to perturbative
techniques. This property implies the presence of two relevant
Hamiltonians for the system: an auxiliary free Hamiltonian $H_0$,
that generates the evolution in the auxiliary linear theory, and a
physical Hamiltonian, $E(H_0)=2(1-e^{-H_0/2})$, which is a
non-linear function of the former \cite{Ashtekar}. The coexistence
of these two Hamiltonians has unexpected physical consequences in
the classical and quantum realms \cite{Ashtekar1, Nosotros}. In
our discussion, we adopt a system of units such that $c=\hbar=
8G_3=1$, where $G_3$ denotes the effective Newton constant per
unit length in the direction of the symmetry axis.

A further step in increasing the complexity of the system, with
the aim at investigating the kind of behavior expected in quantum
gravity, is to consider midi-superspaces that incorporate the
non-linear character of GR without leaving the field theoretic
framework. In this sense, the cylindrically symmetric case with
general polarization is a natural candidate to work with, since it
provides a non-linear generalization of the Eintein-Rosen waves
which consists of two interacting fields \cite{Torre}.

In principle, one might think of quantizing the family of
cylindrical waves with general polarization by taking advantage of
the two Hamiltonians structure that arises in the system. First,
one would employ standard perturbative techniques to analyze and
quantize the auxiliary Hamiltonian. Assuming that this
perturbative quantization were feasible, one would then use the
functional relation between the auxiliary and physical
Hamiltonians in order to construct a quantum theory for the
cylindrical model. Nonetheless, we will see that this quantization
scheme cannot be fully accomplished owing to the
non-renormalizability of the auxiliary theory.

The perturbative quantization of cylindrical gravitational waves
with general polarization has been considered previously by
Niedermaier \cite{Niedermaier}. However, there are some relevant
differences between the approach followed in those references and
ours. Namely, the cylindrical reduction of the gravitational
theory discussed in Ref. \cite{Niedermaier} does not include the
surface term necessary to render the action functionally
differentiable. Actually, this term turns out to supply the
physical Hamiltonian $E(H_0)$ after fixing the time gauge
\cite{AMM}. In this sense, the results obtained by Niedermaier
refer in fact to the auxiliary model of the two Hamiltonians
approach commented above. On the other hand, in the present paper
we follow from the very beginning a path integral approach,
incorporating both a gauge fixing and the corresponding
Fadeev-Popov determinant. This simplifies drastically the
structure of the sigma model attained in Ref. \cite{Niedermaier},
therefore leading to much easier computations. From this
perspective, our discussion can be regarded as a complementary
analysis supporting the arguments of Niedermaier about the
non-renormalizability, modulo a function, of the cylindrical
gravitational waves.

The rest of paper is organized as follows. Section \ref{General}
contains the basic definitions concerning the geometry and
dynamics of the cylindrical waves in vacuo. In that section, we
also derive an expression for the partition function of the
cylindrical gravitational theory, written in terms of the reduced
phase space variables, and identify the auxiliary and physical
Hamiltonians of the system. In Sec. \ref{Perturbative} we discuss
the renormalizability of the auxiliary model. We will show that
the model is non-renormalizable in the standard sense, but belongs
to a wider class of models that are renormalizable modulo a
function, as claimed in Ref. \cite{Niedermaier}. Finally, we
summarize our results and conclude.


\section{\label{General} Cylindrical Waves}

General cylindrical gravitational waves in vacuo are described by
space-times whose metric $\mathbf{g}$ admits a two-parametric,
Abelian, and orthogonally transitive group of isometries. These
space-times are equipped with two linearly independent, spacelike
Killing fields, one of them axial, $X_{(\theta)}$, and the other
translational, $X_{(z)}$. These Killing fields commute,
$[X_{(\theta)},X_{(z)}]=0$, and generate a space-time foliation
orthogonal to the isometry orbits (orthogonal transitivity). Note
that, unlike what happens for the Einstein-Rosen waves, the
Killing vectors are not assumed to be mutually orthogonal,
$\mathbf{g}(X_{(z)},X_{(\theta)})\neq 0$. This allows the presence
of an additional degree of freedom with respect to the linearly
polarized case.

One can always choose coordinates $(t,r,\theta,z)\in
\mathbb{R}\times \mathbb{R}^+\times S^1\times \mathbb{R}$, adapted
to the symmetry of the system, such that the line element takes
the form
\begin{eqnarray*}
ds^2&=&-(N^\bot)^2dt^2+e^{\gamma-\psi}(N^rdt+dr)^2
+R^2e^{-\psi}d\theta^2+e^{\psi}(dz+\phi d\theta)^2\,.
\end{eqnarray*}
The smooth fields $\psi$, $\phi$, $\gamma$, $R$, $N^\bot$ (lapse),
and $N^r$ (radial shift) depend only on the $(t,r)$ coordinates,
satisfy suitable boundary conditions (at the symmetry axis $r=0$
and at spatial infinity $r\rightarrow \infty$) \cite{Torre, MMM},
and parameterize the different vacuum cylindrical metrics. The
scalar fields $\psi$ and $\phi$ have a precise geometrical
meaning, for they are related with the norm and the scalar product
of the Killing vectors,
\begin{eqnarray*}
\psi=\log\big[\mathbf{g}(X_{(z)},X_{(z)})\big]\,,\quad
\phi=\frac{\mathbf{g}(X_{(\theta)},X_{(z)})}
{\mathbf{g}(X_{(z)},X_{(z)})}\,.
\end{eqnarray*}

The dynamics of these cylindrical space-times is inherited from GR
via a symmetry reduction \cite{Torre, AMM}. In this way, one
arrives at the action
\begin{eqnarray} \label{Action}
&&S[N^\bot,N^r,\gamma,R,\psi,\phi,\pi_\gamma,\pi_R,\pi_\psi,\pi_\phi]
=\\
&&\int_{t_1}^{t_2}\left\{\int_0^\infty\bigg[\pi_\gamma\dot{\gamma}
+\pi_R\dot{R}+\pi_\psi\dot{\psi}+\pi_\phi\dot{\phi}-\bigg(N^\bot
\mathcal{H}_\bot+N^r\mathcal{H}_r\bigg)\bigg]dr -
2\bigg(1-e^{-\gamma_\infty/2}\bigg)\right\}dt\,,\nonumber
\end{eqnarray}
where the dot and the prime denote, respectively, the time and
radial partial derivatives. The boundary term involving the value
of $\gamma$ at $r=\infty$, $2\big(1-e^{-\gamma_\infty/2}\big)$,
must be included to render the action functionally differentiable
\cite{MMM}. We have also defined the constraints
\begin{eqnarray*}
\mathcal{H}_\bot\!&:=&
e^{(\psi-\gamma)/2}\bigg[\!-\pi_\gamma\pi_R+2R''-R'\gamma'
+\frac{1}{2R}\left(R^2\psi'^2+\pi^2_\psi
+R^2e^{-2\psi}\pi_\phi^2+e^{2\psi}\phi'^2\right)\bigg]\!\approx
0,\\ \mathcal{H}_r&:=&-2\pi_\gamma'+\pi_\gamma\gamma'+\pi_R
R'+\pi_\psi\psi'+\pi_\phi\phi'\approx 0\,,\end{eqnarray*} which
generate the gauge symmetries of the theory, namely ``bubble" time
evolution and radial diffeomorphisms.


With the aim at developing a quantum theory for the cylindrical
model described by the action (\ref{Action}), we will evaluate the
partition function $\mathcal{Z}$ by means of a path integral
approach. In order to make sense of this path integral, we adopt a
gauge fixing, and introduce the associated Fadeev-Popov
determinant to ensure that the result is independent of the
particular gauge chosen. Employing the usual cylindrical gauge
\cite{Guillermo}
\begin{equation} \label{gf}
R(t,r)=r\,,\quad \pi_\gamma(t,r)=0\,,
\end{equation}
we obtain the expression
\begin{eqnarray*}
\mathcal{Z}&=&\int
[\mathcal{D}\Psi][\mathcal{D}\Pi]\times\prod_{\bar{r}}
\delta\big[\pi_\gamma(\bar{r})\big]
\delta[R(\bar{r})-\bar{r}]\,\det_{\bar{r}}\left(\begin{array}{cc}
\{\pi_\gamma(\bar{r}),\mathcal{H}^\bot\}
&\{\pi_\gamma(\bar{r}),\mathcal{H}_r\}\\
\{R(\bar{r})-\bar{r},\mathcal{H}^\bot\} &
\{R(\bar{r})-\bar{r},\mathcal{H}_r\}
\end{array}\right)\\
&\times&\exp\left\{i\int_{t_1}^{t_2}\int_0^\infty
\bigg[\pi_\gamma\dot{\gamma}+\pi_R\dot{R}
+\pi_\psi\dot{\psi}+\pi_\phi\dot{\phi}-\bigg(N^\bot
\mathcal{H}_\bot+N^r\mathcal{H}_r\bigg)\bigg]dr dt\right\}\\
&\times&
\exp\left\{-i\int_{t_1}^{t_2}2\bigg(1-e^{-\gamma_\infty/2}\bigg)
dt\right\},
\end{eqnarray*}
where
\begin{eqnarray*}
[\mathcal{D}\Psi]&=&[\mathcal{D}N^\bot][\mathcal{D}N^r][
\mathcal{D}\gamma][\mathcal{D}R][\mathcal{D}\psi][
\mathcal{D}\phi]\,,\\ \left[\mathcal{D}\Pi\right]  &=& [
\mathcal{D}\pi_{N^\bot}][\mathcal{D}\pi_{N^r}][
\mathcal{D}\pi_\gamma][\mathcal{D}\pi_R][\mathcal{D}\pi_\psi
][\mathcal{D}\pi_\phi]\,,
\end{eqnarray*}
and
\begin{eqnarray*}
\{R(\bar{r})-\bar{r},\mathcal{H}_r(r)\}&=&R'(r)\delta(\bar{r}-r)\,,\\
\{R(\bar{r})-\bar{r},\mathcal{H}_\bot(r)\}&=&-e^{[\psi(r)-
 \gamma(r)]/2}\pi_\gamma(r)\delta(\bar{r}-r)\,,\\
\{\pi_\gamma(\bar{r}),\mathcal{H}_r(r)\}&=&
-\pi_\gamma(r)\partial_{r}\delta(\bar{r}-r)\,,
\\
 \{\pi_\gamma(\bar{r}),\mathcal{H}_\bot(r)\}
&=& e^{[\psi(r)-\gamma(r)]/2}R'(r)\partial_{r}\delta(\bar{r}-r)+
\frac{1}{2}\mathcal{H}_\bot(r)\delta(\bar{r}-r)\,.
\end{eqnarray*}

The special form of the action and the simplicity of the gauge
fixing term allow us to perform explicitly the integrals on the
non-physical sector. Thus, at the end of the day, we can write
down a formula for $\mathcal{Z}$ exclusively in terms of the
variables of the reduced phase space, avoiding the use of the
standard Fadeev-Popov ghosts. Note first that it is
straightforward to integrate $\pi_\gamma$ and $R$ thanks to the
gauge fixing. One can also easily evaluate the integrals in
$N^\bot$ and $N^r$, which lead to Dirac deltas of the constraints.
By making use of these deltas, it is then possible to integrate
out $\gamma$ and $\pi_R$, attaining an expression for the
partition function (modulo a multiplicative constant
$\mathcal{N}$) which involves only the fields $\psi$ and $\phi$:
\begin{eqnarray}\label{PF}
\mathcal{Z}&=&\mathcal{N}\int [\mathcal{D}\psi][\mathcal{D}\phi]
[\mathcal{D}\pi_\psi] [\mathcal{D}\pi_\phi]\\
&\times&\exp\left\{i\int_{t_1}^{t_2}
\left[\int_0^\infty\bigg(\pi_\psi\dot{\psi}
+\pi_\phi\dot{\phi}\bigg)dr-2\bigg(1-e^{-H_0/2}\bigg)
\right]dt\right\}.\nonumber
\end{eqnarray}
Here, we have defined
\begin{eqnarray*}
H_0&:=&\frac{1}{2}\int_0^\infty\bigg[r\psi'^2
+\frac{\pi^2_\psi}{r}+re^{-2\psi}\pi^2_\phi+
\frac{e^{2\psi}\phi'^2}{r}\bigg]\,dr=\gamma_{\infty}.
\end{eqnarray*}

Some comments are in order. The non-linear character of the
Hamiltonian in the reduced phase space prevents us from performing
the exact integration in the remaining variables. On the other
hand notice that, just like in the linearly polarized case, two
relevant Hamiltonians emerge in the system, as can be read out
from expression (\ref{PF}). Namely, an auxiliary Hamiltonian
$H_0$, that results from the integration of a local density, and
the physical Hamiltonian
\begin{eqnarray*}
E(H_0)&:=&2\,(1-e^{-H_0/2})\,,
\end{eqnarray*}
which is a non-linear function of $H_0$ and, as a consequence,
encodes a non-local dynamics. It is worth emphasizing that, in
contrast to the situation found for linear polarization, $H_0$ is
not the Hamiltonian of a free theory in terms of the scalar fields
$\psi$ and $\phi$, although it reduces to the free Hamiltonian of
the Einstein-Rosen model when the field $\phi$ and its momentum
are switched off (i.e., set equal to zero).

Even though we have followed a path integral approach, equivalent
results about the reduced phase space can be attained in other
different manners; e.g., in a mathematically precise framework, by
making use of standard Hamiltonian techniques \cite{Guillermo},
or, in a rather formal way, by just plugging into the
Dirac-Schr\"{o}dinger action for gravity (equivalent to the
Einstein-Hilbert action up to boundary terms) the gauge fixed line
element
\begin{eqnarray*}
ds^2_{G}=e^{\gamma-\psi}(-dT^2+dr^2)
+r^2e^{-\psi}d\theta^2+e^\psi(dz+\phi d\theta)^2.
\end{eqnarray*}
With this last procedure, one gets
\begin{eqnarray}\label{SigmaAction}
S[ds^2_{G}]=S_\sigma[\psi,\phi]:= \int_{T_1}^{T_2}\int_0^\infty
\frac{r}{2}\left[\left(\dot{\psi}^2-\psi'^2\right)+\frac{e^{2\psi}}
{r^2}\left(\dot{\phi}^2 -\phi'^2\right)\right]drdT,
\end{eqnarray}
an action that leads in fact to the $H_0$ Hamiltonian (owing to
our choice of time $T$).

As we have commented, the non-linear character of $H_0$ poses a
serious obstacle for completing the path integration explicitly.
Nevertheless, one can try to use perturbative techniques in order
to extract relevant physical information. In particular, a natural
way to proceed is to study first the perturbative quantization of
the auxiliary Hamiltonian $H_0$ and, if this task can be achieved,
use the results in a subsequent step to deal with the physical
Hamiltonian $E(H_0)$. In the next section we will explore this
possibility.


\section{\label{Perturbative}Perturbative Quantization of
the Auxiliary Model}

The results of the previous section lead us to consider the
perturbative quantization of the auxiliary theory described by the
action $S_\sigma[\psi,\phi]$, defined in Eq. (\ref{SigmaAction}).
This theory belongs to the class of $x$-dependent sigma models
studied by Osborn in the late 80's \cite{Osborn}.

The $x$-dependent sigma models are theories of maps of the type
$\Phi:X\rightarrow Y$, $x\mapsto \Phi(x)$ between
pseudo-Riemannian manifolds $(X,\eta)$ and $\{(Y,\gamma_x)\}_{
x\in X}$, where the metric of the target manifold $Y$ has an
explicit dependence on the points of the base manifold. The
simplest action for these sigma models is a direct generalization
of the standard one, namely
\begin{eqnarray*}
S_\sigma[\Phi]=\frac{1}{2}\int_X \gamma_{ab}\big[\Phi(x);
x\big]\frac{\partial \Phi^a}{\partial x^\mu}\frac{\partial
\Phi^b}{\partial x^\nu}\, \eta^{\mu\nu}(x)\sqrt{|\eta(x)|}\,dx\,.
\end{eqnarray*}
Particularizing this notation to the $H_0$ theory considered here,
we identify the base manifold as
 $X=\mathbb{R}\times \mathbb{R}^+$, with
coordinates $x=(T,r)$, and metric \cite{footnote}
$$\eta_{TT}=-\eta_{rr}=1.$$
The target manifold $Y=\mathbb{R}^2$ has coordinates
$\Phi=(\psi,\phi)$, and the non-zero metric elements are
\begin{eqnarray*}
\gamma_{\psi\psi}(r)=r\,,\quad
\gamma_{\phi\phi}(\psi;r)=\frac{e^{2\psi}}{r}\,.
\end{eqnarray*}
Therefore, the target manifold $(Y,\gamma_r)$ is a symmetric
space, with a curvature tensor that depends only on $\psi$, apart
from an explicit dependence on the coordinate $r$ of the base
manifold, namely
\begin{eqnarray*}
\mathcal{R}_{abcd}(\psi;r)=\frac{1}{r}
\bigg[\gamma_{ad}(\psi;r)\gamma_{bc}(\psi;r)-
\gamma_{ac}(\psi;r)\gamma_{bd}(\psi;r)\bigg].
\end{eqnarray*}
Without the explicit $r$-dependence, this would be the best
possible theoretical scenario. The dependence on $r$ calls for a
careful study of the perturbative properties of the model and, in
particular, of its renormalizability.

In order to carry out a detailed analysis of the renormalizability
of the action $S_\sigma[\psi,\phi]$, one can employ the standard
dimensional regularization and minimal subtraction scheme
\cite{Osborn}. The bare target metric can be written by means of
the standard pole-loop expansion as
\begin{eqnarray*}
\gamma^B_{ab}(\psi;r)=
\mu^{-\varepsilon}\bigg[\gamma_{ab}(\psi;r)+ \sum_{\nu\geq
1}^\infty\frac{1}{\varepsilon^\nu}\sum_{l\geq
\nu}^\infty\frac{\lambda^l}{(2\pi)^l}\,
T^{(\nu,\,l)}_{ab}\big(\gamma(\psi;r);r\big)\bigg],
\end{eqnarray*}
where $\varepsilon=2-d$ is the dimensional regularization
parameter, $\mu$ is a scale associated with the dimensional
regularization, and $\lambda$ is a loop counter
$(\lambda\propto\hbar)$. The first loop contributions to the bare
metric can be computed by using the formulas in Osborn's paper
\cite{Osborn, foot2}. Thus, the 1-loop and 2-loops contributions
turn out to be, respectively,
\begin{eqnarray*}
T^{(1,1)}_{ab}\big(\gamma(\psi;r);r\big)
&=&\mathcal{R}_{ab}(\psi;r)=-\frac{1}{r}\gamma_{ab}(\psi;r)\,,\\
T^{(1,2)}_{ab}\big(\gamma(\psi;r);r\big)&=&\frac{1}{4}\mathcal{R}_{a
cde}(\psi;r) \mathcal{R}_b\,^{cde}(\psi;r)
=\frac{1}{2r^2}\gamma_{ab}(\psi;r)\,.
\end{eqnarray*}
Actually, owing to the fact that the target manifold is a space of
($r$-dependent) constant curvature, all the loop contributions
take the simple form
\begin{eqnarray*}
T^{(1,l)}_{ab}\big(\gamma(\psi;r);r\big)=\frac{C_l}{r^l}\gamma_{ab}(\psi;r)
\quad (C_l\,\, \mathrm{constant})\,.
\end{eqnarray*}
As a consequence, the renormalized metric of the sigma model can
be written as
\begin{eqnarray*} \gamma^B_{ab}(\psi;r)=\mu^{\varepsilon} \bigg[
1+\frac{1}{\varepsilon}H_1(r)+
\frac{1}{\varepsilon^2}\,H_2(r)+\cdots\bigg]\gamma_{ab}(\psi;r)
=H(r)\gamma_{ab}(\psi;r)\,,
\end{eqnarray*}
where
\begin{eqnarray*}
H_1(r)=-\frac{\lambda}{2\pi r}+\frac{\lambda^2}{8\pi^2r^2}
+O\left(\frac{\lambda^3}{r^3}\right)\,.
\end{eqnarray*}
Explicit formulas for the other functions $H_i(r)$ can be
calculated using Feynman diagrams techniques.

>From the relation $\gamma^B_{ab}(\psi;r)=H(r)\gamma_{ab}(\psi;r)$
between the bare metric and the original one, where $H(r)$ is a
non-constant function, we conclude that the auxiliary $H_0$ model
defined in Eq. (\ref{SigmaAction}) is not renormalizable in the
standard sense. This is just a reflection of the severe problems
that arise in any perturbative approach to the quantization of GR.
The usual perturbative techniques of quantum field theory cannot
be successfully implemented even within the ``simple'' framework
provided by the vacuum cylindrically symmetric waves. On the other
hand, although it is possible to carry out a non-perturbative
quantization of this family of waves \cite{Korotkin}, the physics
behind such a quantization is really difficult to extract, at
least as far as the space-time metric is concerned, and cannot be
interpreted in a standard perturbative way because of the
non-renormalizability of the model.

The kind of non-renormalizability that affects vacuum cylindrical
gravity has been studied by Niedermaier \cite{Niedermaier}. The
key idea is that action (\ref{SigmaAction}) belongs to a wider
class of $r$-dependent sigma models parameterized by an arbitrary
function $h(r)$. In our case, one has to consider the $h$-fa\-mily
of actions
\begin{eqnarray}\label{Ah}
S^h_\sigma[\psi,\phi]=\int_{T_1}^{T_2}\int_0^\infty
\frac{h(r)}{2}\left[\left(\dot{\psi}^2-\psi'^2\right)+
\frac{e^{2\psi}}
{r^2}\left(\dot{\phi}^2-\phi'^2\right)\right]drdT.
\end{eqnarray}
The studied case of cylindrical waves with general polarization
and the gauge fixing (\ref{gf}) would correspond to $h(r)=r$. The
metric of the target space and its Riemann tensor become now
$h$-dependent, namely
\begin{eqnarray*}
\gamma^h_{ab}(\psi;r)&:=& h(r)\gamma^0_{ab}(\psi;r)=
h(r)\left(\begin{array}{cc}1&0\\0&e^{2\psi}/r^2\end{array}\right),\\
\mathcal{R}^h_{abcd}&=& \frac{1}{h}\bigg(
\gamma^h_{ad}\gamma^h_{bc}-\gamma^h_{ac}\gamma^h_{bd}\bigg)\,.
\end{eqnarray*}

Given the special dependence on $h$, action (\ref{Ah}) defines a
maximally symmetric $r$-dependent sigma model. Moreover, the
presence of an adjustable function $h$ in the family of actions
(\ref{Ah}) preserves renormalizability in this broader sense
suggested by Niedermaier and supports the asymptotic safety
scenario. If we compute the bare metric using the formulas of Ref.
\cite{Osborn},
\begin{eqnarray*}
(\gamma^h)^B_{ab}&=&\mu^\varepsilon
\left[1+\frac{1}{\varepsilon}\left(-\frac{\lambda}{2\pi
h}+\frac{\lambda^2}{8\pi^2h^2} +\cdots\right)
+\cdots\right]\gamma^h_{ab}\\
&=&\mu^\varepsilon\left[1+\frac{1}{\varepsilon}
\left(-\frac{\lambda}{2\pi h}
+\frac{\lambda^2}{8\pi^2h^2}+\cdots\right)
+\cdots\right]h\gamma^0_{ab}=
h_B\gamma^0_{ab}=\gamma^{h_B}_{ab}\,\,,
\end{eqnarray*}
we realize that the renormalization of the metric $\gamma^h_{ab}$
amounts to a renormalization of the function $h$, so that we end
up with an element of the same family of actions.

\section{Conclusions}

Parallel to the situation found for Einstein-Rosen waves, we have
seen that two Hamiltonians emerge in the analysis of the general
cylindrically symmetric reduction of GR. The physical Hamiltonian
$E(H_0)$ is a non-linear function of an auxiliary Hamiltonian
$H_0$, defined in terms of a local density. Nonetheless, in
contrast with the relative simplicity of the linearly polarized
case where $H_0$ corresponds to a free scalar field, the auxiliary
Hamiltonian $H_0$ describes now a theory of two interacting
fields. Moreover, this theory is not renormalizable in the usual
sense. Owing to this non-renormalizability, the standard
perturbative approach fails in the attempt to construct a
satisfactory quantum field theory.

Our results agree with those of Niedermaier \cite{Niedermaier},
and in this sense our discussion can be seen as lending further
support to his arguments. In this respect, although the two types
of approaches (Niedermaier's and ours) lead to similar
conclusions, there exist some relevant differences that make this
analysis complementary to a certain extent. A specific feature of
our approach is that we have carefully included the surface terms
needed for the differentiability of the action from the very
beginning. These boundary terms supply (after fixing the time
gauge, with unit lapse at spatial infinity) a physical Hamiltonian
$E(H_0)$ that is a complicated function of the local, auxiliary
Hamiltonian $H_0$. It is only the latter of these Hamiltonians
that can be treated with standard sigma model techniques. In
addition, an advantage of our approach is that, using path
integral methods and Fadeev-Popov determinants, we have been able
to fix the gauge freedom of the system without introducing
ambiguities in the action. The outcome is that the auxiliary
Lagrangian which determines the sigma model acquires a much
simpler form, compared with that considered in Ref.
\cite{Niedermaier}. This considerably simplifies all the
calculations involved. Furthermore, the rationale presented here
seems easy to generalize to other midi-superspace models of
interest, like e.g. the family of Gowdy cosmological solutions
with two polarizations and the spatial topology of a three-torus
\cite{Torre, Mena}. For this family of solutions, one would expect
a situation very similar to that encountered for cylindrical
gravity, except for the interchange of the roles played in the
$x$-dependent sigma model by the time and the spatial coordinate
of the base manifold.

\begin{acknowledgments}
\noindent{This} work was supported by the Spanish MCYT under the
research projects BFM2001-0213 and BFM2002-04031-C02-02. The
authors are grateful to M. Niedermaier for enlightening comments
and to M. Varadarajan for helpful discussions.
\end{acknowledgments}


\begin{thebibliography}{20}
\bibitem{footnote} Actually, owing to the conformal invariance of
$\eta^{\mu\nu}\sqrt{-\eta}$ in two dimensions, the base metric can
be identified only up to a conformal factor.
\bibitem{foot2} These formulas have been widely used in the quantum
field theory literature. Strictly speaking, they are deduced by
computing the determinant of an elliptic operator defined on the
base manifold, and their numerical coefficients may depend on the
topology. Although our base manifold is not compact, we assume
that the standard formulas for the loop-pole coefficients apply as
well to our case.
\end{thebibliography}
\end{document}